# Photoinduced Doughnut-Shaped Nanostructures

[1,2]A. M. Dubrovkin, [1]R. Barillé, [3]E. Ortyl, [3]S. Zielinska,

[1]LUNAM Université, Université d'Angers/UMR CNRS 6200, MOLTECH-Anjou,
2 bd Lavoisier, 49045 Angers, (France).

[2]Division of Physics & Applied Physics, Nanyang Technological University,
50 Nanyang Avenue, Singapore 639798, (Singapore).

[3]Department of Polymer Engineering and Technology, Wroclaw University of Technology,
Faculty of Chemistry, 50-370 Wroclaw, (Poland).

E-mail: regis.barille@univ-angers.fr, alexandre.dubrovkin@univ-angers.fr

We show that an incoherent unpolarized single-beam illumination is able to photoinduce nano-doughnuts on the surface of azopolymer thin films. We demonstrate that individual doughnut-shaped nano-objects as well as clusters of several adjacent nano-doughnuts can be formed and tailored with wide range of typical sizes, thus providing a rich field for applications in nanophotonics and photochemistry.

Azopolymer nanostructures are recognized as an excellent choice for a broad range of fundamental and applied research in modern nanotechnology. Owing to unique photomechanical properties [1 - 2] of azopolymers, these nanostructures show the perfect performance in photoinduced nanopatterning and reshaping by tailored light fields [3 - 4]. Unprecedented flexibility of recently reported photofluidization lithography allows producing well-defined lines, ellipsoids, rectangles and circles on azopolymer surface with structural features of several tenth nanometers [5]. Self-organization is also one possible solution to obtain nanostructures. Micro-droplets of polystyrene solutions on surfaces were used to obtain typical structures like micro-size domes [6]. Other experiments have shown the possibility to create meso-patterning of thin film polymers by controlled dewetting [7]. A 2-D array of femto-litter beakers best described as nano-membrane were investigated by the dewetting of PS film on complex patterned surfaces and exposition to toluene vapor. However the size of the holes is a function of the dimensions of the holes on the underlying substrate. Azopolymer nanostructures are created by illumination with a laser or laser pattern, and an incoherent light from UV-lamp or LED is most frequently used for optical erasing [8]. Another related experiment was performed by illuminating the azobenzene film by a strip-like uniform light pattern formed due to the use of an optical mask [9]. Depending on the polymer architecture either peaks or trenches were observed in the illuminated strips. Laser induced surface deformations were investigated in azobenzene functionalized polymers with one [10] or two-photon isomerization [11] but the obtained patterns were highly dependent of the laser polarization. A good efficiency was obtained with a circular polarization. The width of the deformation induced by the laser was around 2 μm and difficult to overcome with current conventional optics. Moreover two-photon isomerization needs a high power peak laser. Nano-structures were obtained by near-field light generated by an optical fiber probe with a very small aperture diameter of about 50 nm [12] or 120 nm [13]. However the particular set-up for generating near-field light is suitable for evaluation of the response of materials to nano-processing but not for large scale patterns.

Examples of quite rare utilization of incoherent white-light for thin film regular photopattering and reshaping of a single azopolymer nano-sphere has been recently demonstrated [14]. In the last case the size of the nano-objects is sufficiently small to limit the random directions of molecular movements. The use of incoherent light is a promising technique in the case of formation or patterning of nano-objects with dimensions in the order of the coherence property of light. Moreover this technique requires only cheap optical sources. By an exposure of light, azopolymers can produce complex nanostructures such as directed nanoparticle assemblies [15].

In this work we experimentally demonstrate a simple bottom-up approach to form doughnut-shaped nanostructures at the tailored surface of an azopolymer film by an incoherent unpolarized light illumination. The key difference in our approach is the use of an incoherent light for growing nanostructures rather than for optical erasing, and simultaneously directing the final shape of a nanostructure by the initial seed of tiny nanoscale holes.

Doughnut-shaped nano-structures (including nano-doughnuts, nano-rings, nano-toroïds and in particular the example of nano-wells), being a reach object for fundamental study of light localization at nanoscale, have also recently attracted attention due to their efficient use in nano-plasmonics, photochemistry, nano-rectors and sensors. Several successful approaches were developed to produce organic and inorganic doughnut-shaped nanostructures by chemical synthesis and self-assembling [16] as well as colloidal [17] and optical [18] lithography. However they are limited to fabrication of non-reshapeable nanostructures that makes them inefficient for the study of size-dependency of optical and photochemical properties of these structures and their applications to nanoreactors. It remains that a method to produce doughnut-shaped nanostructures with a photoinduced material as azopolymer has not been suggested in the literature. Furthermore all mentioned approaches require relatively complicated fabrication procedures. An example is the fabrication of polymer dot arrays with Electron Beam Lithography [19] that requires a multistep process with dewetting of e-beam exposed film immersed in a mixture of liquids and is considered slow for large area nanopatterning without possibilities to reshape the structures. Polymer thin films are made from a highly photoactive azobenzene derivative containing heterocyclic sulfonamide moieties (ISO1). The details of synthesis of 2-[{4-[(E)-(4-{[(2,6-dimethyl pyrimidin – 4 - yl)amino]sulfonyl}phenyl)diazenyl]phenyl}-(methyl)amino]ethyl2-methylacrylate (fig. 1a) are reported elsewhere [20]. The thin films are prepared by dissolving 75 mg of azopolymer in 1 ml of THF (Tetrahydrofuran), spin-coated on clean glass substrate and let in an oven at 60 °C during one night to eliminate residual solvent. The film thickness was determined with a Dektak profilometer and was around 550 nm. The molecular weight of the polymer determined by GPC was between 14000 and 19000. The glass transition temperature ($T_g$) was 71° C. Figure 1a and 1c shows topography images of the film obtained with different resolution by atomic-force microscope (AFM, Veeco Instruments Inc.) in the tapping mode. The surface of the film is covered with randomly placed nano-holes with a mono-dispersed diameter but with different depths. A typical height cross-section of the film corresponding to dashed lines on AFM images is presented in Figure 1b, d. The topography analysis shows that "big depth" nano-holes have heights of 10-35 nm with full widths at a half minimum (FWHM) of 350-500 nm. "Small depth" nano-holes have heights of 5-10 nm with FWHM of 60-100 nm. The appearance of nano-holes at a polymer film can be achieved by different methods. One is the use of several disturbing factors in the spin-coating procedure such as a non uniform evaporation of a solvent, a presence of small air bubbles and traces of a different solvent (water) in the polymer solution [21]. These methods are difficult to control and the final result is non reproducible. Here we address the issue to create nano-holes easily reproducible with a monosize diameter on a surface of an azopolymer film. The method chosen is a solvent-droplet-induced-dewetting of thin azopolymer films on glass substrates. The contact of a thin polymer film to a solvent droplet reduces the glass transition temperature to below the room temperature as the solvent molecules penetrate into the film matrix. The stable polymer thin film is destabilized by the introduction of polar interactions. More informations on the surface topology observed with spontaneous hole development in thin polymeric films can be found elsewhere [22]. A 3 μl droplet of THF (Tetrahydrofuran) is dropped off on the thin film surface. This solvent is the one used to dissolve our azopolymer powder. THF has a low boiling point of 66 °C and an evaporating rate of 2.3. It is a medium evaporating solvent compared to other solvents. The quantity of THF entering into the polymer film is gradual. The high molecular weight azopolymer film was stable after the experiment. A small quantity of THF entering in the azopolymer thin film is not sufficient to induce enough chain mobility and the glassy state is maintained. The kinetics of hole growth is very fast and difficult to evaluate. A simple rough qualitative evaluation has shown that the radii of holes increase linearly with time during the hole growth. The length scale period $\lambda_f$ between surface structures is a function of the film thickness as well as the surface and interfacial tensions of the film and the substrate material. We calculate the mean length scale of the period between nano-holes created by dewetting of droplets of THF on the surface of our azopolymer thin films. The volume of droplets is 3 μl. We consider in the calculation the material as pure PMMA which is the largest part of the matrix material. We suppose that the consideration of azobenzene chromophore into the calculation will not induce a significant modification of the material molecular weight. We find an average value of $\lambda_f$ of 1.23 ± 0.1 μm in agreement with the theoretical results of 1.16 μm given by [23]:

$$\lambda_f = \left(\frac{M_e \gamma}{|S|}\right)^{1/3} \frac{h_o^{7/6}}{M^{1/2}}$$

where M is the polymer weight of PMMA, $h_o$ the film thickness (800 nm), $M_e$ the molecular weight between entanglements (7000 Da), $|S|$ the spreading coefficient (70 MPa) and $\gamma$ the surface tension (-0.8 mN/m). We note that the dimension of the rims is also monosize. Holes opening during the dewetting process occur at locations having the smallest thickness and can be caused by fluctuations of the film thickness. The topographic measurement with an AFM have revealed an average fluctuation of the azopolymer thin film surface of 3 ± 1 nm. The location of holes appears randomly but can be controlled with initial changes of the film topography by dewetting on already prepared patterned surfaces [24].

Owing to the small depth of nano-holes, prepared films do not exhibit a strong light scattering and appear as a flat colored glass for an unaided eye. The thickness of the film presented in Figure 1 (measured with AFM scanning of the scratch made in the film till the glass substrate) is approximately 500 nm. By means of the highly photoactive azopolymer material property, these films reveal high efficiency of photoinduced mass transport and self organization under the single beam laser illumination, resulting in nanoscale surface grating formation [25]. From the other hand, as a result of a different light redistribution between a coherent and incoherent illumination due to diffraction or scattering, incoherent light can indifferently direct the photopatterning process in comparison with a laser illumination. We choose this technique to uniformly modify the holes formed on the surface and induce a polymer growth due to mass transport. This approach with a white source was chosen for the creation of patterns by photolithography of azopolymer liquid crystal thin films through masks [26]. The initial surface modification with holes acts as mask for the photofluidization of these nano-structures.

In the experiment (Figure 2a) we used incoherent white light from xenon halogen lamp to initiate photoinduced mass transport in the film. Effective light power affecting the sample in the azopolymer absorption band was 225 mW/cm$^2$. In order to prove the ability of photoinduced nano-structuring we made a detailed scanning electron microscopy (SEM) of the film after illumination during 30 minutes. Residual solvent in the holes are totally eliminated after light exposure. Figure 2b shows typical SEM image of the film. Numbers of randomly placed doughnut-shaped nano-objects and spherical protrusions are clearly distinguished. For better visualization of a single nano-doughnut we made the close-up SEM image of the film at 60° tilt (Figure 2c). Nano-doughnuts represent a structure which consists of a central hole and a surrounding ring. We define the diameter of a doughnut as a diameter of the ring top part. Detailed analysis shows that average diameter of photoinduced doughnuts is 450 ± 50 nm. Most frequently the distance between a nano-doughnut and neighboring nano-objects is greater or comparable to its diameter after illumination. This property of the film makes it possible to work with nano-doughnuts in some applications as with isolated objects. The sample was illuminated during 30 minutes. The volume of the doughnut has grown to reach a mean value of 0.065 μm$^3$ meaning a growth rate of 0.0022 μm$^3$/min if we consider that 30 minutes is the time to reach the saturation of the phenomenon. During this illumination time the nano-object growth can be considered evolving linearly before saturation. The force F needed to exert an elastic deformation on the nano-object is derived from Hertz theory [27] which considers the contact deformation of elastic spheres under normal loads in the absence of adhesion and friction:

$$F = \frac{4}{3} G D_o^{0.5} \Delta D^{1.5}$$

where G is the elastic modulus of PMMA (1.8 – 3.1 GPa), $\Delta D = D_o - D$ the deformation with $D_o$ and D the initial and final height. We find an elastic force of 7.8 ± 1 μN acting on the hole to produce a nano-doughnut.

Although the most large area of the illuminated film contains typical distribution of nano-objects presented in Figure 2b,c, we observed regions with a different topology close to the edge of the sample. As a result of a spin-coating procedure the edge area of the film after fabrication has non uniform morphology which changes abruptly. Figure 2d-g shows two typical cases of nano-structuring after illumination in the edge area. SEM images of Figure 2d demonstrate a possibility of fabricating comparable densities of nano-doughnuts with smaller (gray color of the central hole) and greater (black color of the central hole) heights. More detailed SEM studies along with additional AFM measurements showed a possibility to fabricate nano-doughnuts with various depths up to the substrate level. The second case (Figure 2e-g) represents photoinduced growing of nano-doughnuts with smaller lateral sizes. We observed doughnuts with diameters varying from 150 nm to 400 nm. Compare with a typical film topography (Figure 2b) one can benefit from this variation in any possible applications which need a simultaneous utilization of similar nano-objects with different sizes.

We calculate the radius of gyration ($R_g$) of the nano-doughnut. This is the average squared distance of any point in the object (polymer toroïd) from its center of mass. The parameter $R_g$ of a toroïd is a function of the radius R and the ring width (2r) [28] and is given by:

$$R_g = R\left(1 + \frac{3}{4}Z^2\right)^{0.5}$$

with Z = R/r. We take R = 0.6 μm and 2r = 0.25 μm from AFM measurements. The value of $R_g$ is stable for all the nano-doughnuts. The small discrepancies between the measurements of $R_g$, are attributed to the fact that the ring-shaped structures obtained by dewetting have a uniform toroïdal structure instead of core-shell nano-ring.

Quite rare but several regularly neighboring doughnuts or a combination of a doughnut and a spherical protrusion appear with the distance smaller than its diameter (Figure 2h,i). This regularity can assist any future applications which need a doughnut-to-doughnut coupling effects.

In order to show a height distribution of photoinduced nano-doughnuts, a typical film topography (far from the edge) was studied using AFM (in contact mode). The AFM image (Figure 3a) shows the picture of randomly placed nano-objects similar to SEM measurements. The height information is presented using cross-sections along dashed lines across several nano-doughnuts. Resulting profiles are plotted in the graph (fig. 3b). By taking into account the film thickness of approximately 500 nm it is clear that some doughnut-shaped nano-objects are grown at the surface of the film while others are induced through the whole film thickness until the substrate. Several measurements for different nano-doughnuts have shown that typical heights of a doughnut rim above the film surface vary from 50 nm to 150 nm. All presented information leads to the natural classification of observed doughnut-shaped structures as nano-doughnuts and nano-wells. Nano-doughnuts represent structures with a central hole lying near the film surface while nano-wells have greater depth. The example of a close-up single nano-doughnut topography obtained by AFM is presented in the Figure 3c. Corresponding cross-section along the dashed line is plotted in figure 3d.

In summary, we have experimentally demonstrated the fabrication of doughnut-shaped nanostructures with a wide range of heights and diameters. Self-organized behavior of doughnuts formation in the azopolymer film represents a simple way of nano-structuring. Furthermore, due to the wide topography variation of the azopolymer film, nano-doughnuts can be used both as single nano-objects and as clusters of several nano-doughnuts for performing doughnut-to-doughnut coupling effects. In the future the choice of the solvent or mixture of solvents will help to optimize the control of the interfacial tension and effective destabilizing forces leading to different patterns. Finally, we emphasize that a sufficiently wavelength-dependent nature of an azopolymer response to light pushes forward the utilization of the doughnuts as functional nano-objects and nanoreactors.

Acknowledgements

The authors would like to acknowledge the financial support of CNRS and Région Pays de la Loire. We are grateful to SCIAM laboratory at the University of Angers for the help with obtaining SEM images.

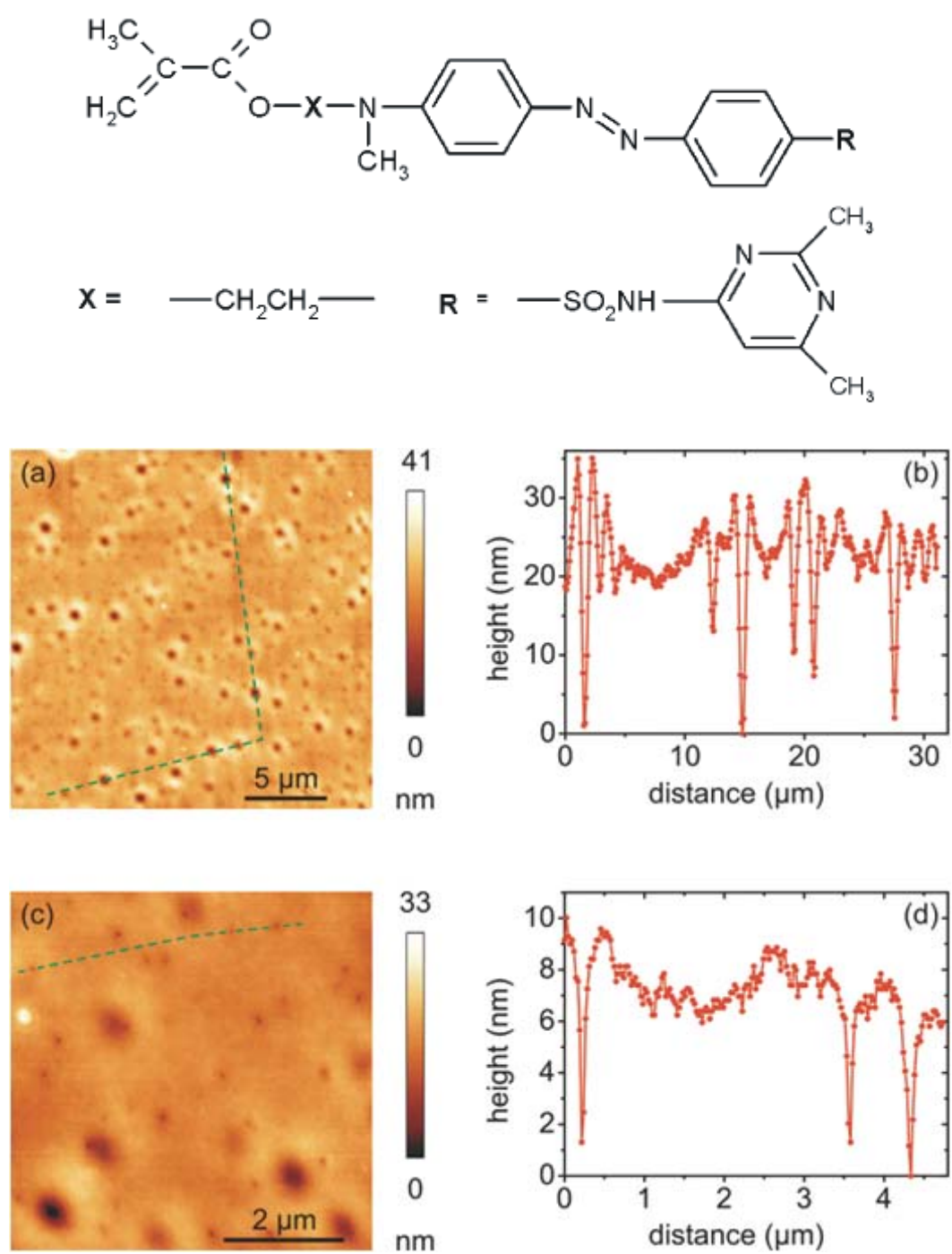

**Figure 1.** Chemical structure of azopolymer thin film and AFM characterization of the surface. (a, b) A typical topography and corresponding height cross-section shown by the dotted line. (c, d) Zoom of the topography obtained by AFM scanning in the centre part of the region presented in the image (a) and corresponding cross-section.

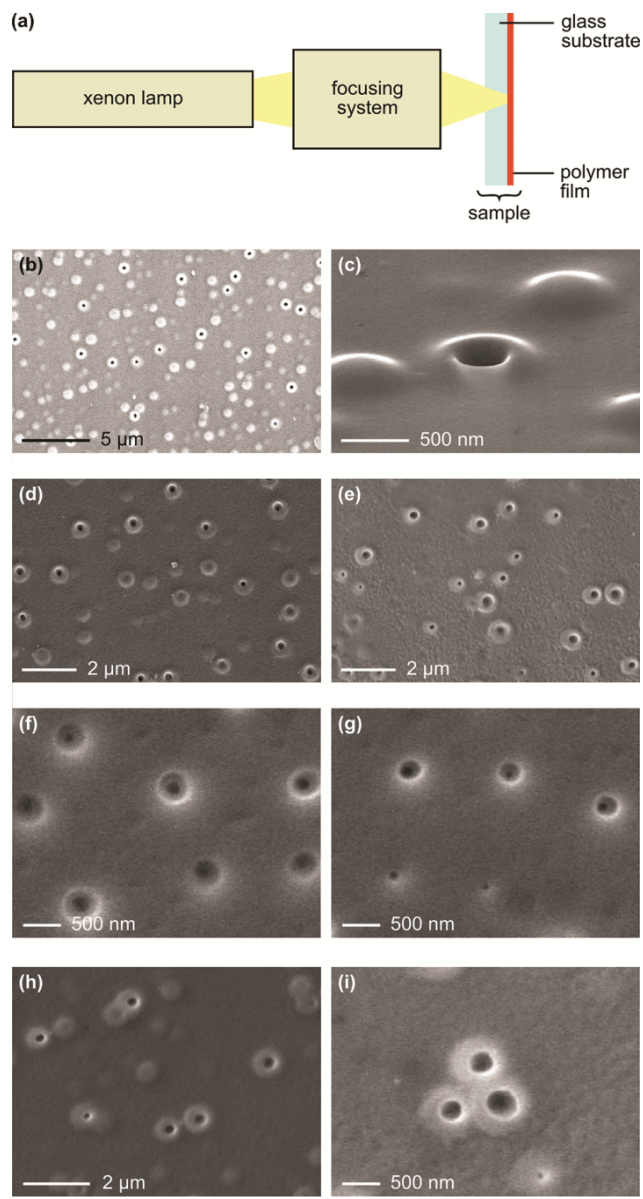

**Figure 2.** (a) A scheme of the white light illumination of the sample in the experiment. (b-i) SEM images of different regions of the azopolymer film after the illumination. A tilt of the sample at the image (c) is 60°. Images (b, d - i) were recorded without tilting.

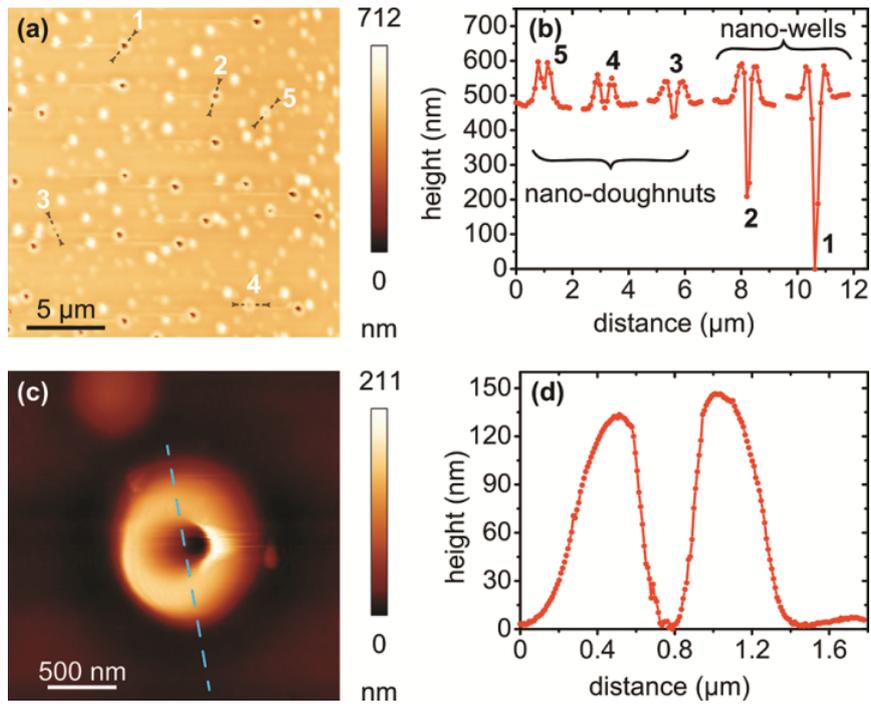

**Figure 3.** Topographical images of the azopolymer film after the white light illumination. (a, b) A typical large scale topography obtained by AFM and corresponding height cross-section over five marked lines. (c, d) A typical close-up topographical image of azopolymer nano-doughnut obtained by AFM and corresponding height cross-section over the dotted line.